\documentclass[12pt,fleqn]{article}
\usepackage{latexsym}

\begin{document}
\title{Quantum Error Correction in the Zeno Regime}
\author{Noam Erez$^{a}$, Yakir Aharonov$^{b,c}$, Benni Reznik$^{b}$ and Lev Vaidman$^b$
 {\ }\\
\small a) {\em \small Institute for Quantum Studies and Department of Physics, Texas A\&M University,}\\
{\em \small College Station, TX 77843-4242, USA} \\ 
\small b) {\em \small School of Physics and Astronomy, Tel Aviv
University, Tel Aviv 69978, Israel}\\ \small c) {\em \small Department
of Physics, University of South Carolina, Columbia, SC 29208, USA}
}
%\date{October 16, 2001}
\maketitle

\begin{abstract}
In order to reduce errors, error correction codes (ECCs) need to
be implemented fast. They can correct the errors corresponding to
the first few orders in the Taylor expansion of the Hamiltonian of
the interaction with the environment. If implemented fast enough,
the zeroth order error predominates and the dominant effect is of
error prevention by measurement (Zeno Effect) rather than
correction. In this ``Zeno Regime'', codes with less redundancy
are sufficient for protection. We describe such a simple scheme,
which uses two ``noiseless'' qubits to protect a large number, $n$, of information
qubits from noise from the environment. The ``noisless qubits'' can be realized by treating
them as logical qubits to be encoded by one of the previously introduced encoding schemes.

%which encodes $n$ logical qubits in $n+10$ physical qubits, and
%compare it with ECCs and Dynamical Decoupling (DD).
\end{abstract}

\section{Introduction}

Quantum error correction schemes based on encoding, consist of the following steps: 1. encoding $n$ logical qubits into $m>n$ physical ones,
(2. introduction of errors) 3. syndrome measurement - projecting the state onto one of a number of subspaces corresponding to a discrete set of errors
4. error correction\cite{Shor1}. For a review on various error correction codes see Ref.\cite{KLABVZ}

The error correction capabilities of the various codes are usually described in terms of the discrete set of errors in step 3: a code is said
to correct a discrete set of errors with certainty. This makes the treatment very
similar to the theory of algebraic codes of classical Information Theory.

In most of the ECCs that have been proposed, the discrete set of errors to be corrected consists of a Pauli operators acting on only a few qubits.
Why is this error set interesting? If one assumes that the different qubits are located at well separated physical locations (as in an ion-trap realization
of a quantum computer), then it is reasonable to assume that the environmentally induced errors for the different qubits are independent. If the correction is
implemented fast enough, then the dominant contribution will come from the first few orders in the Taylor expansion of the interaction with the environment,
and these are spanned by the discrete set.

It is important to note, that one of the syndromes corresponds to
``no-error'', and that this corresponds to the zeroth order error.
Therefore, implementing steps 1-3 often enough will also have the
effect of reducing the error\cite{LevZeno}. In other words, Measuring the
syndrome often enough prevents errors. This is the Quantum Zeno Effect\cite{Zeno} (QZE).
(for the syndrome degree of freedom!): repeatedly making a
projective measurement can freeze the dynamics. For a discussion of the implementability of the 
QZE, as well as the inverse effect, see Ref\cite{Kurizki}
%The first to
%suggest using the Zeno Effect to suppress errors was W.
%Zurek\cite{Zurek} (back in 1984!). Since then, a number of  
A number of Quantum Codes utilizing the error prevention that occurs in the Zeno
limit have been proposed \cite{Zurek, Zeno2, LevZeno}. 

In the ``Zeno Regime'', the ``first order'' error correction codes
are overly redundant. It has been shown in Ref.\cite{LevZeno} that in
that regime, the 5-qubit code\cite{5qubit} can be replaced by a 2-qubit
encoding per logical qubit (for an even number of logical qubits). In other words, 
%with Shor's code 
error correction codes can protect $n$ logical qubits which are encoded in $5n$ physical qubits\cite{5qubit}, and previous error prevention codes (Zeno) 
can achieve this for encoding in  $2n$ physical qubits. 

In the next section we describe a ``Zeno'' error correction code which encodes $n$ logical qubits in $n+4$. This consists of two encoding
steps. The first step consists of encoding into $n+2$ qubits for a code that protects against single qubit errors that can occur in only a 
definite set of $n$ qubits. The second step is to encode the two unprotected qubits with the 4-qubit code of Ref\cite{LevZeno}.

Dynamical Decoupling\cite{VL} schemes of EC also bear some similarity to the Zeno Effect, but differ in important ways.
Recently, a unification of Dynamical Decoupling and the QZE has been suggested\cite{Lidar}

\section{A Zeno Error Prevention Code}

In close analogy to the linear codes of classical Information Theory, 
a quantum code which encodes $n$ logical qubits into $m=n+r$ physical qubits, is defined to be a unitary linear mapping
$C:\mathcal{H}_{logical}\mapsto \mathcal{H}_{physical}$ where $\dim (\mathcal{H}_{physical})=2^m$ and $\dim (\mathcal{H}_{physical})=2^n$.
This must be implemented by a unitary acting on a Hilbert space which includes $\mathcal{H}_{logical}$ and $\mathcal{H}_{physical}$ as subspaces.
Nothing essential is lost if we assume that this large Hilbert space is simply $\mathcal{H}_{physical}$. This just means that the logical qubits
are initially stored in a $2^m$ dimensional subspace of $\mathcal{H}_{physical}$. Then $\mathcal{H}_{physical}$ is isomorphic to the product 
$\mathcal{H}_{system}\otimes \mathcal{H}_{ancilla}$ where $\mathcal{H}_{system}$ is isomorphic to $\mathcal{H}_{logical}$ and 
$\mathcal{H}_{ancilla}$ is a dimensional space we will call the ``ancilla''.
Now to fix the identity of the subspace $\mathcal{H}_{logical}$, we have to choose a one-dimensional subspace of $\mathcal{H}_{ancilla}$, denoted by
$\textrm{Span}\{|in\rangle_{ancilla}\}$ (where `in' stands for initial state), so $\mathcal{H}_{logical}=\mathcal{H}_{sys}\otimes|in\rangle_{anc}$.
To summarize, we rewrite the encoding operation: ``state of $m$ logical qubits is mapped into state of $n$ physical qubits'', 
as ``$|\psi\rangle_{sys}\otimes|in\rangle_{anc}$ is mapped to encoded state of same space''.
Formally then, the encoding operation corresponds to an operator $C$ 

\begin{equation}
C:\mathcal{H}_{sys}\otimes\mathcal{H}_{anc} \mapsto \mathcal{H}_{sys}\otimes\mathcal{H}_{anc}.
\end{equation}

%A Quantum Code which encrypts $n$ logical qubits in $m=n+r$
%physical ones, is a unitary mapping $C:\mathcal{H}_{n}\otimes
%|in\rangle \mapsto \mathcal{H}_m$, where $|in\rangle \in
%\mathcal{H}_r$ is a known initial state. In any physical application, our encoding, $C$, like any
%other operation has to be implemented by some unitary operator acting on the entire system, consisting 
%of the one in which the original ``logical'' qubits are encoded as well as all the other qubits. 
%Since $\mathcal{H}_{n+m}$ is isomorphic to $\mathcal{H}_n\otimes\mathcal{H}_m$, we may without loss of generality
%identify $C$ with this unitary operation. Let us call the Hilbert space of the original ``logical'' qubits, $\mathcal{H}_{n}$, 
%the \emph{system} space, $\mathcal{H}_{sys}$; and the auxiliary space, the space of the \emph{ancilla}: $\mathcal{H}_{anc}$.
%The encoding operator $C$ then maps $\mathcal{H}_{sys}\otimes\mathcal{H}_{anc}$ into itself:
%\begin{equation}
%C:\mathcal{H}_{sys}\otimes\mathcal{H}_{anc} \mapsto \mathcal{H}_{sys}\otimes\mathcal{H}_{anc}
%\end{equation}

After this encoding step,
some interaction with the environment introduces
noise\footnote{Note that in a more realistic model, we would have
to consider noise during all the other steps as well, which would
require a fault tolerant approach}. The most general Hamiltonian
describing an environment (a system with state Hilbert space,
$\mathcal{H}_{env}$) interacting \emph{independently} with each qubit is:

\begin{equation} H_{noise} = -\mathcal{E}\sum_{\stackrel{i=1,...,n}{b=0,...,3} }
\sigma_{b}^{s_i}\otimes A_{env}^{i,b},\end{equation}
%then the time evolution operator is (with $\mathcal{T}$ denoting time ordering), 
%$U_{noise}(t)=\mathcal{T}\exp{(-i\int_0^t H(s) ds/\hbar )}$. For short times,
%$t=\varepsilon \hbar/\mathcal{E}$, we can expand $N(\varepsilon)\equiv U_{noise}(\varepsilon\hbar/\mathcal{E})$:
then the time evolution for short times, $t=\varepsilon \hbar/\mathcal{E}$, is:
\begin{equation} N(\varepsilon) \equiv U_{noise}(\varepsilon\hbar/\mathcal{E}) = 1 + i\varepsilon\sum_{i, b}
\sigma_b^{s_i}\otimes A_{env}^{i,b} +O(\varepsilon^2) \label{eq:Taylor}
\end{equation}

We claim that it suffices to use a 2-qubit ancilla (i.e. $\dim \mathcal{H}_{anc}=4$) to protect an unknown state in $\mathcal{H}_{sys}$ from 
the first order terms in the expansion of the noise operator, Eq. \ref{eq:Taylor}. Namely, 
for a certain choice of the initial state of the ancilla, $|in\rangle_{anc}$, 
and encoding operator\footnote{We make here two implicit assumptions about the environment. 
Firstly, we assume (in common with most works on error correction) that it is initially in a product state with our ancilla and system. Secondly
we assume its initial state to be pure. The latter assumption is not really needed, but was made in order to simplify the notation. If the scheme 
works for any pure state, by linearity it will work for any mixture as well. The first assumption is more subtle. },  $C$, which will be specified below

\begin{equation}CNC|in\rangle|\psi\rangle|0\rangle_{env} =
|in\rangle_{anc}|\psi\rangle_s|\phi\rangle_{env}+ \varepsilon |\bot
\rangle +O(\varepsilon^2),\end{equation}where $|\bot\rangle \in
|in\rangle_{anc}^{\bot}\otimes\mathcal{H}_{sys}\otimes
\mathcal{H}_{env}$. While it is to be expected that $|\bot\rangle$ should be orthogonal to the initial overall state,
it should be emphasized that \emph{it is orthogonal to the initial state of the ancilla}.
This means that a projective measurement on the ancilla alone, will
find it in the initial state $|in\rangle$ with probability
$1-O(\varepsilon^2)$, and in this case the system will be in it's
(unknown) initial state, $|\psi\rangle$ (also with probability
$1-O(\varepsilon^2)$). The $O(\varepsilon^2)$ probability not to find the system in state $|\psi\rangle$ even for a favorable
outcome of the measurement of the ancilla, corresponds to the $O(\varepsilon^2)$ terms in Eq. \ref{eq:Taylor}, i.e. to higher order errors
($n$ qubit errors, $n>1$).

Let us show this for a 2-qubit ancilla, \emph{assuming the noise acts only on the other} $n$ \emph{qubits}.
 The logical qubits reside originally in $|\psi\rangle \in \mathcal{H}_{sys}$
($\dim\mathcal{H}_{sys}=n $). Let us choose some arbitrary basis
$\{|a\rangle\}_{a=0}^3$ for $\mathcal{H}_{anc}$, and choose
\mbox{$|in\rangle_{anc} = \frac{1}{2}\sum_{a=0}^{3}|a\rangle$}, and
define our encoding operator, $C$, to be:
\begin{equation} C=\sum_{a=0}^3(|a\rangle\langle a|)\otimes
\prod_{i=1}^n\sigma_a^{i},\label{defC}\end{equation}where
$\sigma_a^{i}$ acts on the $i$th qubit in $\mathcal{H}_{sys}$.
Then the encrypted $n+2$ qubit state is

\begin{equation} C|in\rangle_{anc}|\psi\rangle_{sys} = \frac{1}{2}\sum_{a=0}^3|a\rangle (\otimes_{i=1}^n\sigma_a^{i} )|\psi\rangle. \end{equation}

Then for $b\neq 0$,

\begin{eqnarray}  C\sigma_b^jC = \sum_{a,i,i'}|a\rangle\langle
a|(\otimes_i\sigma_a^i)\sigma_b^j(\otimes_{i'}\sigma_a^{i'}) = \nonumber \\
 \sum_a |a\rangle\langle a|\sigma_a^j\sigma_b^j\sigma_a^j =
\left(\sum_a|a\rangle\langle
a|c_a^{(b)}\right)\sigma_b^j
\end{eqnarray} where $c_a^{(b)}= \left\{
\begin{array}{l} +1\ a=0,b \\-1 \displaystyle{\ otherwise}
\end{array} \right.$. \\ For $b=0$, we have the trivial result $c^{(0)}_a=1$ for all $a$. For all $b$ then

\begin{equation} C\sigma_b^j C|in\rangle|\psi\rangle =
\left(\frac{1}{2}\sum c_a^{(b)}|a\rangle
\right)\sigma_b^j|\psi\rangle \equiv |\tilde{b}\rangle
\sigma_b^j|\psi\rangle,
\end{equation} the $\{ |\tilde{b}\rangle \}_{b=0,...,3}$ form an orthonormal
basis, and $|\tilde{0}\rangle=|in\rangle$\footnote{It is not hard
to see that if the original basis $\{ |a\rangle \}_{a=0,...,3}$ is
chosen as the simultaneous eigenfunctions of $\sigma_z\otimes 1,
1\otimes \sigma_z$, the $\{ |\tilde{b} \rangle \}_{b=0,...,3}$
basis consists of the simultaneous eigenfunctions of
$\sigma_x\otimes 1, 1\otimes \sigma_x$.}.

%\newpage

Finally, we can write the effect of encoding on the
environmentally induced errors (we assume, as before, that initially the
system is in a product state with the environment, and that the
latter is in a pure state, $|0\rangle_{env}$):

\begin{eqnarray} CNC|in\rangle_{anc}|\psi\rangle_{sys}|0\rangle_{env} = \nonumber &  \\ |\tilde{0}\rangle_{anc}|\psi\rangle_{sys}|\phi\rangle_{env}
+i\varepsilon\sum_{\stackrel{b\neq 0}{j=1,...,m}}
|\tilde{b}\rangle_{env} \sigma_b^j|\psi\rangle_{s} A^{j,b}|0\rangle_{env}
+O(\varepsilon^2)= \nonumber & \\
 |in\rangle_{anc}|\psi\rangle_s|\phi\rangle_{env}+ \varepsilon |\bot \rangle +O(\varepsilon^2) & \end{eqnarray} 

So far we have, somewhat artificially, considered the case where we
have at our disposal $2$ ``privileged'' qubits which are exempt from
noise (the $2$ qubits we had singled out as our ancilla). To
remedy this, we note that these ancillary qubits can be each
encoded in $5$ ``ordinary'' physical qubits subject to the same
noise as the others, using the famous $5$-qubit code\cite{5qubit}. 
Alternatively, we can encode the two special qubits in 4 physical qubits using the scheme of Ref.\cite{LevZeno}
 to get an ``all Zeno'' code.

\newpage

\section{Two Interpretations}

In this section we will mainly restrict the discussion to the case
of one logical qubit ($n=1$) for the sake of simplicity. The generalization to the case of arbitrary $n$ is straightforward, and
will be discussed very briefly.

\subsection{Heisenberg Representation}

In the previous sections, we have concentrated on the
Schr\"{o}dinger representation to make the treatment more easily
comparable to the usual error correction schemes. However, a few
authors have also looked at error correction in the Heisenberg
representation, see for example\cite{Gottesman}. Our original
derivation was in the latter representation, and is perhaps
somewhat more natural. In the Heisenberg representation, the
encoding and decoding operators, which act before and after the
``noise operator'' respectively, are seen as acting on the latter operator.
The choice of initial state of the ancilla, and the projection
onto the syndrome subspaces (also defined by the state of the
ancilla in our scheme) are seen as a pre- and post-selection steps.
The desired effect of the encoding, decoding and postselection 
should be that when the postselection of the desired subspace succeeds, the
effective noise operator becomes trivial as far as the system is concerned.

To be more concrete, let us write $N$ for the noise operator which acts on our system and the environment:
\[N:\mathcal{H}_{sys}\otimes \mathcal{H}_{env}\mapsto \mathcal{H}_{sys}\otimes \mathcal{H}_{env},\]$U_{enc},\ U_{dec}$ for the encoding and decoding operators (respectively)
acting on the system and ancilla:
\[U_{enc,\ dec}: \mathcal{H}_{sys}\otimes \mathcal{H}_{anc} \mapsto \mathcal{H}_{sys}\otimes \mathcal{H}_{anc} \] and $|in\rangle$ and $|fin\rangle$ for the preselected 
(resp. postselected)
states of the ancilla, then  
\begin{equation} _{anc}\langle out |U_{dec}NU_{enc}|in \rangle_{anc}
\propto 1_{sys} \label{eq:defin}
\end{equation}

Before we write the explicit form of these objects in our scheme, let us motivate it with a simple example that works for
a single qubit ``system'' a single qubit ancilla and a noise operator of the special form $N=O_0^{env}+O_1^{env}\sigma_y^{sys}+O_2^{env}\sigma_z^{sys}$ 
(i.e. not involving $\sigma_x^{sys}$). For the usual definition of the CNOT (=conditional flip) operator:

\[C_x=(|0\rangle\langle 0|)_{ancilla}+(|1\rangle\langle 1|)_{ancilla}\sigma_x^{sys} \]($|0\rangle \equiv |s_z=+\hbar/2\rangle$, $|1\rangle\equiv |s_z=-\hbar/2\rangle $),

we have the following property:

\begin{equation} C_x\sigma_y^{sys}C_x = \sigma_y^{sys}\sigma_z^{anc}, ~C_x\sigma_z^{sys}C_x = \sigma_z^{sys}\sigma_z^{anc} \label{firsteq}
\end{equation}and so

\[ _{anc}\langle \uparrow_x|C N C|\uparrow_x \rangle_{anc} = O_0^{env}
\]which is Eq.(\ref{eq:defin}) with $U_{enc}=U_{dec}=C$ and $|in\rangle_{ancilla} =|fin\rangle_{ancilla}=|\uparrow_x\rangle$.

In order to generalize this treatment to treat general errors (for a one qubit system), we will need two ``conditional-flip'' operators, where
the ``flip'' denotes a Pauli operator: $C_a^{1,2}=|0 \rangle_{1}\langle 0|\otimes 1_2
+|1\rangle_{1}\langle1|\otimes\sigma_a^2 (\ a=x,y,z$) 
%is a CNOT
%operator (in an appropriate basis) on the space of two qubits:
 which acts on $\mathcal{H}_1\otimes \mathcal{H}_2$.
Writing $\mathcal{H}_{anc}=\mathcal{H}_1\otimes\mathcal{H}_2$, $\mathcal{H}_{sys}=\mathcal{H}_3$, and
%$|0\rangle=|\uparrow\rangle_z,\ |1\rangle=|\downarrow\rangle_z$,
%as usual; and 
$C=C_y^{1,3}C_z^{2,3}$ (regular operator product), then Eq.(\ref{firsteq}) generalizes to:
%is an operator on $\mathcal{H}^{1,2}\otimes \mathcal{H}^{3}\equiv
%\mathcal{H}_{anc}\otimes \mathcal{H}_{sys}$. Then we claim:
\begin{eqnarray} C(1_{anc}\otimes \sigma_x)C^\dag & = &(\sigma_z\otimes \sigma_z)^{anc}\otimes (\sigma_x)^{sys} ,\ 
\nonumber \\  C(1_{anc}\otimes \sigma_y)C^\dag & = &(1 \otimes \sigma_z)^{anc}\otimes (\sigma_y)^{sys},\
\nonumber \\  C(1_{anc}\otimes \sigma_z)C^\dag& = &(\sigma_z\otimes 1)^{anc}\otimes (\sigma_x)^{sys}
\end{eqnarray}which follow from:

\begin{eqnarray} C^{1,2}_a\sigma_b^2 C^{1,2}_a & = & (|0\rangle\langle 0|)_1\sigma^2_b+(|1\rangle\langle 1|)_1(\sigma_a\sigma_b\sigma_a)_2 = \nonumber \\
& = &  [|0\rangle\langle 0| + |1\rangle\langle 1|(-1)^{\delta_{ab}+1}]_1 \sigma_b^2 =
 \left\{ {  1\otimes\sigma_b \ a=b \atop \sigma_z\otimes\sigma_b \ a\neq b } \right. .
\end{eqnarray}Note also that:

\begin{eqnarray} C=(|0,0\rangle\langle 0,0|)_{anc}1_{sys}+(|0,1\rangle\langle 0,1|)_{anc}\sigma_z^{sys}+ \\ \nonumber
(|1,0\rangle\langle 1,0|)_{anc}\sigma_y^{sys}
+i(|1,1\rangle\langle 1,1|)_{anc}\sigma_x^{sys} \end{eqnarray}
which is almost our definition eq(\ref{defC}) for $n=1$.

Let us introduce for these expressions the more compact notation:

\begin{equation} C\sigma_a^{sys}C^{\dag}= \Sigma^{anc}_a \sigma_a^{sys}
\end{equation}and add $\Sigma_0 = 1$.

With the definition, $|in\rangle = |\uparrow_x,\ \uparrow_x\rangle$, and the fact that $_{anc}\langle in|\Sigma_a|in\rangle_{anc}=\delta_{a0}$, we finally have:

\begin{eqnarray}
_{anc}\langle in|CUC^{\dag}| in \rangle_{anc}& \simeq &  _{anc}\langle in| e^{i\varepsilon\sum_{i,a}H^{env}_{i,a}\Sigma^{anc, s_i}_a\sigma^{s_i}_a }|in\rangle_{anc}
  \simeq e^{i\varepsilon\sum H_{i,a} \langle in|\Sigma^{anc, s_i}_a|in\rangle \sigma^{s_i}_a } \nonumber \\ & = & e^{i\varepsilon \sum_i H^{env}_{i,0}}\otimes 1^{sys}
\end{eqnarray}

This last result can be interpreted as follows: for an ancilla initially in state $|in\rangle$, and postselected to be in the same position, the operator $U$ is
effectively reduced to a harmless one acting on the environment alone. In this form, it is very reminiscent of the dynamical error correction. Nevertheless, this approach has the advantage, that unlike dynamical error correction, it does not require that the noise be slowly varying, only that it be ``small''.

%\footnote{The expression in the exponent is actually Aharonov's ``weak value'' of $\Sigma$ with respect to }

\subsection{Error Prevention as Teleportation in Time}

Our scheme can also be written in yet another way, related to
measurement theory, if we look at the controlled-not operation as
a measurement of the state of the system qubit.

It was shown in \cite{LevTel} that ``crossed nonlocal measurements'' performed on two separated qubits 
%Fig.1a
\begin{equation}
  \label{eq:cnm}
  \Bigl({\sigma_1}_x(t_1) - {\sigma_2}_x(t_1 +\epsilon)\Bigr){\rm mod4}, ~~~~
\Bigl({\sigma_1}_y(t_1 +\epsilon) - {\sigma_2}_y(t_1)\Bigr){\rm
mod4}
\end{equation}
yield two-way teleportation or ``swapping'' of the states of the qubits. The four possible outcomes of these measurements define the correction which 
have to be performed on the qubits to complete the teleportation. The particular results
\begin{equation}
 \label{eq:cnm1}
  \Bigl({\sigma_1}_x(t_1) - {\sigma_2}_x(t_1  +\epsilon)\Bigr){\rm mod4} =
\Bigl({\sigma_1}_y(t_1 +\epsilon) - {\sigma_2}_y(t_1)\Bigr){\rm
mod4} =0
\end{equation}
correspond to immediate teleportation without need for corrections. In
this teleportation procedure, the times of the interaction with the
second qubit can be changed provided that the order remains the same
\begin{equation}
  \label{eq:cnm2}
  \Bigl({\sigma_1}_x(t_1) - {\sigma_2}_x(t_2 +\epsilon)\Bigr){\rm mod4}, ~~~~
\Bigl({\sigma_1}_y(t_1 +\epsilon) - {\sigma_2}_y(t_2)\Bigr){\rm
mod4}
\end{equation}

In particular, we can arrange that $ t_2 > t_1 +\epsilon$.
%, see Fig. 2b. 
The ``identity'' of the second particle is not important, the procedure
teleports the state of the qubit to any particle with which the
interactions are performed. Thus, we can make the interactions at times
$ t_2$ and $t_2  +\epsilon$ with the particle which had the first qubit. In
this case we teleport the quantum state of a particle to the particle
itself, but at a later time: teleportation in time!
%, see Fig. 1c.: teleportation in time!

In fact, teleportation in time (as well as teleportation to another
particle at time-like interval) is much easier to perform than
teleportation to a space-like interval. We have to perform the
following two-time measurements \cite{AAD}
\begin{eqnarray}
 \label{eq:2tm}
  \Bigl({\sigma}_x(t_1) - {\sigma}_x(t_2 +\epsilon)\Bigr){\rm mod4},\nonumber \\   
\Bigl({\sigma}_y(t_1 +\epsilon) - {\sigma}_y(t_2)\Bigr){\rm
mod4} 
\end{eqnarray}
where $ t_1< t_2 <t_1'< t_2'$.  This measurements are much easier to
perform than measurements of nonlocal variables required
for the two-way teleportation. There is no need to have entangled
particles in the measuring device. A single qubit replaces the
entangled pair. The coupling to the qubit is the same as the coupling
to the entangled qubits of the pair and, in fact, it is just CNOT in
the appropriate basis, exactly the same interaction which was used in
the procedure described in the previous section.

The measurements (\ref{eq:cnm}) and, in general, the measurements (\ref{eq:2tm}) might
have four possible outcomes. However, if our system was not disturbed
between $ t_1$ and $ t_2'$ (except for measurements (\ref{eq:2tm})),
only a single outcome is possible:
\begin{equation}
\label{eq:2tm0}
   \Bigl({\sigma}_x(t_1) - {\sigma}_x(t_2 +\epsilon)\Bigr){\rm mod4}=   
\Bigl({\sigma}_y(t_1 +\epsilon) - {\sigma}_y(t_2)\Bigr){\rm
mod4}=0
\end{equation}
This is the outcome which  corresponds  to the teleportation without corrections.
Indeed, the measurements  (\ref{eq:2tm}) are also verification measurements of the 
particular type of a two-time state \cite{VPhd}
\begin{equation}
 \label{eq:2ts}
\Psi_{t_1,t_2} = {1\over {\sqrt 2}}(  \langle{{\uparrow}}|_{t_1}
|{\uparrow}\rangle_{t_2} + \langle{{\downarrow}}|_{t_1}
|{\downarrow}\rangle_{t_2}
\end{equation}
which is generated by vanishing Hamiltonian at the time period $[t_1,
t_2]$. If, during this period there will be a small disturbance then
the measurement, due to Zeno effect will, with high probability, still
have the outcome (\ref{eq:2tm0}) and it also will nullify the action of
the disturbance, i.e., prevent errors during this time.

For discussion of the general case of protection $N$ qubits we have to
look more closely on the process of measurement of the two
time-variables (\ref{eq:2tm}). These measurements require two qubits 
prepared in a particular state before time $t_1$ which undergo two CNOT
(conditional flip) interactions. One qubit in the $\sigma_x$ basis at
times $t_1$ and $t_2 +\epsilon$ and another, in the $\sigma_y$ basis at
times $t_1 +\epsilon$ and $t_2$. The measurements, after the
interaction with the system which verify  that the test qubits have
not changed their state complete the measurement. If the system was not
disturbed, then the coupling of the test qubits with the system does not
prevent the test qubits to verify the absence of disturbing of
another qubit using coupling corresponding to two-time measurements performed on
another qubit
\begin{eqnarray}
%\label{eq:2tm}
  \Bigl({\sigma'}_x(t_1-\delta) - {\sigma}_x(t_2 +\epsilon +\delta)\Bigr){\rm mod4},\\   
\Bigl({\sigma}_y(t_1 +\epsilon-\delta) - {\sigma}_y(t_2+\delta)\Bigr){\rm
mod4}
\end{eqnarray}
If both systems are under small disturbance (such that probability of
flipping of both qubits is negligible) then the procedure: preparation
of the test qubits coupling of the test qubits with the two systems,
and final verification that the the test qubits have not changed their
state will lead, through Zeno effect to prevention of errors in the
two systems.
The general case is treated as before.

\begin{center}
ACKNOWLEDGEMENTS
\end{center}

We would like to acknowledge the support of the Basic Research Foundation of
the Israeli Academy of Sciences and Humanities. Y.A. acknowledges the support of the National Science Foundation.

%The later method
%complicates significantly the procedure, since it is difficult to
%arrange coupling of the encoded test qubit with the qubits it is going
%to protect.

\end{document}